\documentclass[11pt]{article}
\pdfoutput=1
\usepackage{jheppub}
\usepackage{graphicx,float,flowchart}
\usepackage{amssymb}
\usepackage{subcaption}
\usepackage{amsmath}
\usepackage{slashed}
\usepackage{hyperref}
\usepackage{caption}
\usepackage{xcolor}
\usepackage{dsfont}
\usepackage{verbatim}
\usepackage{setspace}
\usepackage{slashed}
\usepackage{xcolor}

\newcommand\beq{\begin{equation}}
\newcommand\eeq{\end{equation}}

\def\shrug{\texttt{\raisebox{0.75em}{\char`\_}\char`\\\char`\_\kern-0.5ex(\kern-0.25ex\raisebox{0.25ex}{\rotatebox{45}{\raisebox{-.75ex}"\kern-1.5ex\rotatebox{-90})}}\kern-0.5ex)\kern-0.5ex\char`\_/\raisebox{0.75em}{\char`\_}}}

\title{Entropy, Entanglement and Swampland Bounds in DS/dS}

\preprint{\today}

\author[a]{Hao Geng,}
\author[a,b]{ Sebastian Grieninger,}
\author[a]{ and Andreas Karch}
\affiliation[a]{Department of Physics, University of Washington, Seattle, Wa, 98195-1560, USA}
\affiliation[b]{Theoretisch-Physikalisches Institut, Friedrich-Schiller University of Jena, Max-Wien-Platz 1, 07743 Jena, Germany}
\emailAdd{hg666@uw.edu, sebastian.grieninger@gmail.com, akarch@uw.edu}

\abstract{We calculate the entanglement entropy of the de-Sitter (dS) static patch in the context of the DS/dS correspondence. Interestingly, we find that there exists a one parameter family of bulk minimal surfaces that all have the same area. Two of them have appeared earlier in the literature. All of them correctly calculate the dS entropy. One surface yields the entanglement between the two different CFTs that provide the holographic dual of the bulk DS geometry. The second surface describes the entanglement across the horizon in the boundary static patch. The other surfaces describe a mixture of these two concepts. We also show that in the presence of extra matter fields the former entanglement entropy always exceeds the dS entropy. We interpret  this result in the context of entropy bounds in de Sitter space and the swampland program.}

\begin{document}
\maketitle

\section{Introduction}
While holography provides a definition of quantum gravity in asymptotically anti-de Sitter (AdS) space, de Sitter (dS) is still a mystery. The question of quantum dS gravity has seen renewed interest recently due to the bold dS swampland conjectures put forward starting with the work of \cite{Obied:2018sgi}. Despite the large evidence for a landscape of dS vacua in string theory \cite{Bousso:2000xa}, for example in the context of the KKLT construction \cite{Kachru:2003aw}, it was asserted that consistency of quantum gravity in dS requires scalar potentials with either a non-vanishing derivative, completely destroying dS to begin with, or a large negative mass squared \cite{Garg:2018reu,Ooguri:2018wrx}, rendering dS unstable. Stable dS quantum gravity was banned to the swampland. Clearly it would be desirable to check some of these assertions within a well defined framework of quantum gravity of dS space.

One holographic framework for such a consistent description of quantum dS gravity is provided by the DS/dS correspondence \cite{Alishahiha:2004md}. Based on the observation of \cite{Karch:2003em} that dS space can be viewed as a Randall Sundrum type \cite{Randall:1999vf} setup with two asymptotic AdS regions joined by a UV brane with a localized graviton, \cite{Alishahiha:2004md} proposes that the dual description of dS$_D$ space consists of two $D-1$ dimensional conformal field theories (CFTs), with a UV cutoff, living on dS$_{D-1}$. They are coupled to each other and to $D-1$ dimensional gravity. As we will review in detail, the geometry of dS-sliced dS-space very naturally suggests this interpretation. Two asymptotic regions realize a near horizon geometry which is identical to that of dS-sliced AdS space. Latter is known to describe the infrared (IR) of a conformal field theory living on dS$_{D-1}$. The full dS$_D$ geometry glues together these two asymptotic regions along a UV slice which traps a dynamical graviton by the standard Randall-Sundrum mechanism \cite{Randall:1999vf}.  Steady progress over the last several years \cite{Dong:2018cuv,Gorbenko:2018oov,Dong:2010pm} has put the proposal on a more solid foundation. Maybe most noteworthy, the construction of \cite{Gorbenko:2018oov} allows one, in principle, to start with a CFT dual to AdS gravity and deform it into one dual to dS gravity.

In this work we want to study entanglement entropy in the context of the DS/dS correspondence using the standard holographic RT prescription \cite{Ryu:2006bv}. This idea is not new. In particular, \cite{Dong:2018cuv} constructed minimal area surfaces that are interpreted as giving the entanglement between the left (L) and right (R) CFTs: by tracing over the degrees of freedom of one, one obtains a density matrix for the other. Quite remarkably, this calculation gives exactly the de Sitter entropy of the $D$-dimensional de Sitter space, an apparent success of this framework. There is a second RT surface that is also well known from the context of Randall Sundrum gravity: if we study the $D-1$ dimensional de Sitter space on the {\it UV slice} in its static patch, we can calculate an entanglement entropy associated with tracing out all degrees of freedom (L and R) behind the horizon. This completely reproduces the entropy of the $D-1$ dimensional de Sitter space on the slice (which is the same as the full $D$ dimensional de Sitter entropy to begin with). This is an example of a statement that is true in general \cite{Emparan:2006ni,Myers:2013lva}: in any Randall Sundrum setup, the gravity on the boundary is induced and, in this case, one expects that the Bekenstein-Hawking entropy associated with any horizon, including the cosmological dS horizon\footnote{This fact was also noted in \cite{Nguyen:2017ggc}.}, is completely accounted for by entanglement entropy across the horizon. Somewhat surprisingly, we find a one parameter family of solutions that interpolate between those two extreme cases, and they {\it all} yield the same entropy. There are apparently an infinite number of different ways to reproduce the entropy of de Sitter space from entanglement.

This richness of equivalent RT surfaces gives us an opportunity to explore potential consistency requirements (``swampland criteria") on matter we can add to dS gravity. Turning on these matter fields, which is supposed to correspond to turning on extra allowed couplings in the dual CFTs, we deform dS$_D$ space to a generic warped spacetime with dS$_d$ slices. We find that, consistent with the results of \cite{Emparan:2006ni,Myers:2013lva}, the entanglement entropy across the horizon on the slice continues to reproduce exactly the de Sitter entropy for arbitrary deformations. This is however no longer true for any of the other RT surfaces. Since the de Sitter entropy is believed to be the largest possible entropy in de Sitter space, we can ask under what conditions the other entanglement entropies end up {\it no larger} than the de Sitter entropy. Interestingly, we find that {\it any} reasonable matter, that is as long as it obeys the Null Energy Condition, leads to a L/R entanglement entropy that is always larger than the de Sitter entropy on the UV slice. Potential interpretations of this result will be discussed in the concluding section.

Our paper is organized as follows: in the next section, section \ref{setup}, we will review the basic setup of the DS/dS correspondence and establish our notation. We study the various RT surfaces in DS as well as their interpretations in section \ref{rt}. In section \ref{general} we turn to a general warpfactor and establish general inequalities obeyed by the various entanglement entropies. We speculate about potential interpretations in section \ref{conclude}.

\section{Setup and notation}
\label{setup}
Let us set up the notation for the basic DS/dS framework \cite{Alishahiha:2004md}.
We study a $D$ dimensional de-Sitter space (which we will refer to as DS$_D$), with a holographic dual living on a $d=D-1$ dimensional de-Sitter space (which we will refer to as dS$_d$). The holographic dual involves two CFTs with large central charge coupled to dynamical gravity. We refer to the Newton's constant of gravity on DS$_D$ as $G$ and to the one of the dynamically induced graviton on dS$_d$ as $g$. In the spirit of the DS/dS correspondence the metric on dS$_D$ is written as
\beq
\label{ds}
ds^2_{DS} = dr^2 + e^{2 A(r)} ds^2_{dS}
\eeq
with
\beq
e^{A(r)} = L \cos \left ( \frac{r}{L} \right ) .\label{asympdS}
\eeq
Here $L$ is the curvature radius of DS$_D$ and $ds^2_{dS}$ is the metric on a unit dS$_d$. The warp factor $A(r)$ interpolates between the horizons at $r/L = \pm \pi/2$. It's maximum is at $r=0$, the ``UV slice", at which the wavefunction of the localized graviton also peaks. In what follows we will also be interested in slightly more general $A(r)$ with the same asymptotic behavior at $r = r_\text{min}$ and $r=r_\text{max}$ and a maximum warpfactor at $r=r_\text{m}$ with $A'(r_\text{m})=0$ and $A''(r_\text{m})<0$.
The full metric of the dS$_d$ the holographic dual lives on is given by the warpfactor at $r=0$ times this unit metric, that is $L^2 ds^2_{dS}$ for the case of DS$_D$. The $D$-dimensional Newton constant $G$ and the $d$ dimensional Newton constant $g$ are related by the properly weighted volume of the internal space
\beq \label{genericnewton} g^{-1} = G^{-1} \, \int dr\, e^{(D-3)A}.\eeq
For the case of DS$_D$ this yields
\beq
g^{-1}= G^{-1}L^{D-3}\!  \int_{-\frac{\pi L}{2}}^{\frac{\pi L}{2}} dr\, \cos^{D-3} \left(\frac{r}{L}\right)
= G^{-1}L^{D-2}\! \int_{-\frac{\pi}{2}}^{\frac{\pi}{2}} dy\, \cos^{D-3}(y)
= G^{-1}L^{D-2} \, \sqrt{\pi}\, \frac{ \Gamma(\frac{D}{2}-1)}{\Gamma(\frac{D-1}{2} ) }.\label{dsEE}
\eeq
To specify entangling surfaces we also need to commit to a coordinate system on the slice.
For the unit metric dS$_d$ we chose static coordinates
\beq
\label{patch}
ds^2_{dS} = -(1-\rho^2) \, d \tau^2 + \frac{d \rho^2}{1-\rho^2} + \rho^2 d\Omega^2_{D-3} = - \sin^2 \beta d \tau^2 + d \beta^2 + \cos^2 \beta \, d \Omega^2_{D-3}
\eeq
where we employ two different parametrizations of the radial coordinate related by
\beq
\rho = \cos (\beta).
\eeq
A single static patch is parameterized by values of $\rho$ with $0 \leq \rho \leq 1$, that is
$0 \leq  \beta \leq \frac{\pi}{2}$. The origin maps to $\beta=\pi/2$. The surface $\rho=\rho_h=1$ ($\beta = 0$) is the cosmological dS$_d$ horizon. Its area is $A_{D-3}$, the area of the unit $D-3$ sphere. Correspondingly it is associated with an entropy
\beq S_d = \frac{L^{D-3} A_{D-3}}{4 g}. \eeq
Note that the area of a unit $D-2$ and $D-3$ sphere are related by\footnote{To see this, note that
we can write the metric on a unit $D-2$ sphere are
$$d\Omega_{D-2}^2 = d \theta^2 + \cos^2 \theta \, d \Omega_{D-3}^2.$$}
\beq A_{D-2} = A_{D-3} \int_{-\frac{\pi}{2}}^{\frac{\pi}{2}} d\theta \, cos^{D-3} \theta .\eeq
This is exactly the same integral relating the Newton's constants in \eqref{genericnewton}. So, as already noted in \cite{Karch:2003em}, we have
\beq S_{d} = \frac{L^{D-3} A_{D-3}}{4 g} = \frac{L^{D-2} A_{D-2}}{4 G} = S_{D}. \eeq
That is, the area of the DS$_D$ horizon in $D$-dimensional Planck units is equal to the area of the dS$_d$ horizon in $d$-dimensional Planck units; holography correctly calculates the de-Sitter entropy.

Last but not least, let us introduce one more coordinate system, the DS$_D$ static patch. Here the metric can be written as
\beq
\label{static}
ds^2_{DS} = -H(R) dT^2 + \frac{dR^2}{H(R)} + R^2 d\Omega_{D-2}^2
\eeq
with
\beq
H(R) = 1 - \frac{R^2}{L^2}
\eeq
and
\beq
d\Omega_{D-2}^2 = d \theta^2 + \cos^2 \theta \, d \Omega_{D-3}^2
\eeq
as in the footnote above with $-\pi/2 \leq \theta \leq \pi/2$. The two DS$_D$ metrics from \eqref{ds} and \eqref{static} are related by
\beq R = L \sqrt{\rho^2 + \sin^2(\frac{r}{L}) (1 - \rho^2)}, \quad \quad \sin \theta = \frac{\cos(r/L) \rho}
{\sqrt{\rho^2 + \sin^2(\frac{r}{L}) (1 - \rho^2)}}.\eeq
We will not make much use of this other than to note that the DS$_D$ horizon $R=L$ maps to $\rho=1$ for all $r$. That is, the surface defined by the union of all the dS$_d$ horizons on each slice is the dS$_D$ horizon.

\section{Entanglement Entropies}
\label{rt}

In order to shed some light on the de Sitter entropy, we would like to calculate entanglement entropies associated with spherical entangling surfaces centered around the origin of the static patch. That is, we chose our entangling surface in the metric \eqref{patch}  to be given by
\beq \tau=0, \quad \quad \rho=\rho_0 \eeq
with $0 < \rho_0 \leq 1$. The special case $\rho_0=1$ corresponds to the dS$_d$ horizon.
Our task is to find RT surfaces associated with these entangling surfaces.

\subsection{Spatial entanglement}

One class of surfaces (class U) one can look for are of the form $\rho(r)$, standard ``U"-shaped surfaces hanging down towards the IR. Since the DS geometry extends to both sides, the RT surfaces will have to be double sided. The Lagrangian for such RT minimal surfaces is easiest written in terms of $\beta(r)$ and reads
\beq
{\cal L}_{I} = L^{D-3} \cos^{D-3}(\beta) \cos^{D-3} \left ( \frac{r}{L} \right ) \sqrt{1 +L^2\, \cos^2 \left ( \frac{r}{L} \right )
 (\beta')^2}.
\eeq
It is easy to check that for $\rho_0=1$ there is a simple solution: $\beta(r) = 0$, that is
\beq \rho(r)=1. \eeq
All terms in the equation of motion vanish identically either due to the fact that $\rho'=0$ or that the derivative of $\cos \beta$ vanishes at $\beta=0$.
This U-shaped RT surface is just the DS$_D$ horizon! The associated entanglement entropy is completely accounting for the de-Sitter entropy:
\beq S_{U} = S_{dS}. \label{eq:classUEE}\eeq
This appears to be a very appealing picture.

Of course this is not the only U-shaped entangling surface one can find. The $\rho \equiv 1$ solution we just constructed smoothly caps of at $r^*/L= \pi/2$. We can similarly solve for U-shaped entangling surfaces that cap of at $0 \leq r^*/ L \leq \pi/2 $. Naively one would have expected these to correspond to entangling surfaces associated with $\rho_0 <1$. As we will see, this turns out not to be the case. In fact all these entangling surfaces end at the horizon on the UV slice, $\rho_0=1$, and they all have the same area. But before we demonstrate this fact, let us first take a look at the second special entangling surface.

\subsection{Integrating out one CFT}

A second class, class D, of RT surfaces has been constructed in \cite{Dong:2018cuv} and these exist and are smooth for {\it any} $\rho_0$. This time we parameterize the surface by $r(\rho)$ and observe that $r=0$ is a solution simply due to the fact that the warpfactor has a maximum at $r=0$. In detail, the Lagrangian in terms of $r(\beta)$ this time reads
\beq
{\cal L}_{II} =L^{D-3}\cos^{D-3}(\beta) \cos^{D-3} \left ( \frac{r}{L} \right ) \sqrt{(r')^2 +L^2\, \cos^2 \left ( \frac{r}{L} \right )
}.
\eeq
For $r(\beta) \equiv 0$ all terms in the equations of motion vanish again identically either by the fact that $r'=0$ or that $A'(0)=0$.
The RT surface associated to generic $\rho_0$ completely lives on the UV slice and is given by the volume enclosed by the $\rho=\rho_0$ surface. For the special case of $\rho_0=1$, the horizon as the entangling surface, this volume is given by
\beq
V_{D-2} = A_{D-3} \int_{0}^{\pi/2} d\beta \, \cos^{D-3} (\beta)\,.
\eeq
This is once again the {\it same} integrand as in the relation of the Newton constants in \eqref{dsEE}. One important difference however is the range of the integration. In the previous two instances the range was from $-\pi/2$ to $\pi/2$, we integrated the cosine from minimum to maximum back to minimum. Here we only integrate from the maximum at $\beta=0$ (the horizon)  to the minimum at $\beta=\pi/2$ (the origin of the static patch). Correspondingly
\beq 
\label{staticee}
S_{D, \text{static}} = \frac{1}{2} S_{dS}. \eeq

The interpretation given for these class D surfaces in \cite{Dong:2018cuv} was that they correspond to tracing out the degrees of freedom of one of the two CFTs. This makes intuitive sense since the entangling surface separates the left and right half of the warped spacetime. How precisely this is defined when, as in the case considered here, we only do this in a bounded spatial region has been left open in \cite{Dong:2018cuv}. But it is very easy to understand these surfaces in the case that one traces out one of the two CFTs in the entire spatial part of the dS$_d$ spacetime. In this case the RT surface simply is the entire spatial part of dS$_d$ localized at $r=0$. The spatial volume of dS$_d$ is the volume of a unit-sphere, while the volume inside the static patch was half of this unit sphere. Correspondingly, this ``global" entanglement entropy is twice what we got in the static patch calculation. We find in complete agreement with \cite{Dong:2018cuv} that\footnote{Of course we would also get this entropy if we were to take two copies of the class D static patch RT surfaces in order to once again get a "two sided" RT surface.}
\beq
\label{globalee}
S_{D,\text{global}} = S_{DS}.
\eeq
That is, we recovered the de Sitter entropy from entanglement entropy in two very different ways. In the first calculation we studied the entanglement entropy across the horizon in the holographic dual and found that the corresponding entanglement entropy accounts for the full dS entropy. In the second calculation, one traced over one of the two CFTs in an entire spatial slice. In the case of a simple DS$_D$ spacetime both calculations give the same correct answer. We will see in section \ref{general} that apparently this is no longer the case once we allow ourselves to turn on deformations.

\subsection{A one parameter family of entangling surfaces}

In this subsection we wish to construct the general smooth U-shaped RT surface.
Starting from the Langrangian
\begin{equation}
\mathcal L_I=L^{D-3}\,\rho(r)^{D-3}\,\cos\left(\frac rL\right)^{D-3}\ \sqrt{1+\frac{L^2\,\cos^2\left(\frac rL\right)}{1-\rho(r)^2}\ \rho'(r)^2},
\end{equation}
we may derive the equations of motions by a variational principle.

\subsubsection{Numerical construction of entangling surface in $D=5$}
In the following discussion we restrict ourselves to $D=5$ (and set $L=1$). Focusing on $0\leq r\leq \pi/2$ and $0\leq \rho\leq 1$, we are able to determine the right half of the minimal surfaces which smoothly end on $r^\star$ by solving the equations of motion numerically. The left half is identical and simply yields an overall factor of 2. The solutions are uniquely parametrized by $r^*$; a second potential integration constant at $r^*$ is eliminated by requiring regularity. At $r=0$ one can extract $\rho(0)$ and $\rho'(0)$. One would expect that both of these integration constants depend non-trivially on $r^*$. Instead we find a class of solutions which determines the two free integration constants to $\rho(0)=\rho_0=1$ and $\rho'(0)=0$.

All of these solutions have the same boundary value and first derivative but their second derivatives differ depending on $r^\star$. 
The second derivative approaches zero for $r^\star\to\frac\pi2$ and diverges for $r^\star\to0$. In order to determine the area, we have to integrate the Lagrangian
\begin{equation}
    A=2\,\int_{0}^{r^\star}dr\, \mathcal L_I[\rho(r)],
\end{equation}
where $\rho(r)$ is our numerical solution. Interestingly we find for all the solutions with different $r^\star$ the same area, $\pi/2$.
This matches exactly eq. \eqref{eq:classUEE}, which means the associated entanglement entropy is the same as the de-Sitter entropy (and the entanglement entropy associated to the $\rho\equiv1$ solution).

We can identify the special case of $r^*=\pi/2$ as our class U solution from before. The case $r^* \rightarrow 0$ turns into (2 copies) of class D. Note that class D was defined to be a single disk shaped region localized at $r=0$, whereas here we get one from the left and one from the right. The generic $r^*$ surfaces smoothly interpolate between these two cases.

\subsubsection{Analytical solution for the entangling surface in $D$ dimensions}
Starting from the observation that the entanglement entropy seems to be independent of $r^\star$ we were wondering if we can construct an analytical solution for the entangling surface. Since we know that $\rho(r)\equiv 1$ solves the equation of motion and all our numerically constructed solutions vanish at $r^\star$, we may calculate pertubative corrections to these known solutions. By matching the pertubative expansions, we find that the $D$ dimensional entangling surface is given by
\beq
\label{analyticsolution}
\rho(r)=\pm \sqrt{1-\tan\left(r/L\right)^2/\tan\left(r^\star/L\right)^2}\quad\Leftrightarrow\quad\beta(r)=\pm\arcsin\left[\tan\left(r/L\right)/\tan\left(r^\star/L\right)\right].
\eeq
\begin{figure}[h]
    \centering
   \includegraphics[width=7.1cm]{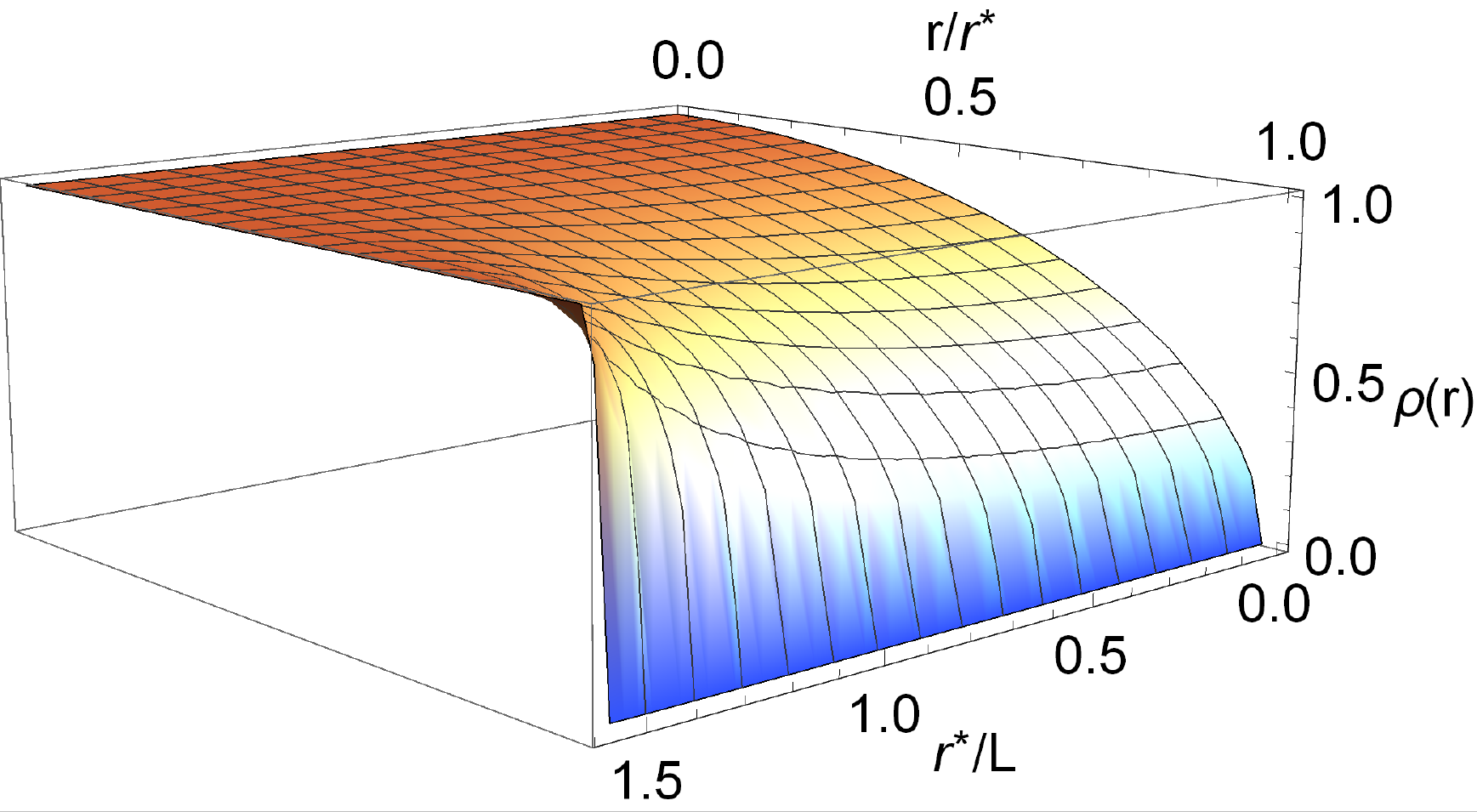}
        \caption{Solution for $\rho(r)$ in dependence of $r^\star$.}\label{pic:rhof}
\end{figure}
By plugging this analytical solution in the Lagrangian and integrating over $r$, it is straightforward to show that the entangling surfaces reproduce the correct entanglement entropy eq. \eqref{dsEE}.

The exact interpretation of these extra entangling surfaces is not clear. The two extreme cases were argued to be integrating out only the left degrees of freedom for class D, versus equally integrating out L and R for class U. Presumably the interpolating solutions describe some mix of integrating out L and R. While we are agnostic about the detailed definition in the field theory, we will take the point of view that they are all calculating slightly {\it different} ways of tracing out degrees of freedom. So each one of them is a valid entanglement entropy calculation in the CFT. 

Given that we didn't find any smooth class U solutions giving us a $\rho_0 <1$, one may wonder what happened to these boundary conditions. Shooting from $r=0$ with $\rho(0)\neq 1$ leads to solutions which ``blow up" in the bulk (i.e. are not reaching zero). Even if $\rho'(0)$ is set negative, $\rho(r)$ quickly turns around and shoots back out towards $\rho=1$. Apparently this is not a question in the dual theory we are supposed to ask, except for in the case of the class D surface. As noted above, latter does allow for $\rho_0 <1$. This may potentially be due to the fact that the holographic dual description involves gravity. So calculating entanglement entropy for generic coordinate surfaces may not be a well defined quantity. We will have to leave a more direct field theory understanding of this phenomenon for the future.
\subsubsection{Geometric Interpretation}
\begin{figure}[H]
    \centering
   \includegraphics[width=7.1cm]{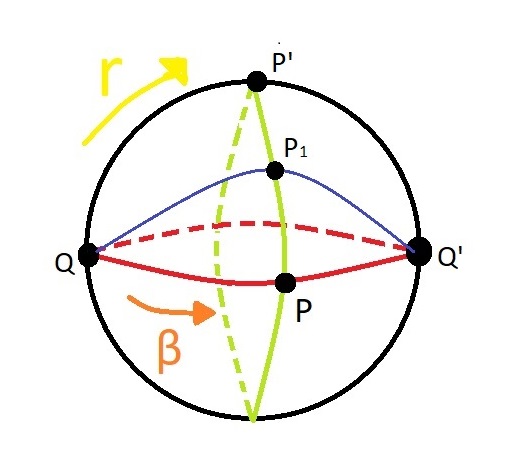}
        \caption{Geometric Interpretation of the analytical solution eq. \eqref{analyticsolution}.}\label{pic:GeomInter}
\end{figure}

From \eqref{patch} we can see that the spatial geometry is spherical where codimension-1 extremal surfaces are grand spheres. For the sake of convenience, we provide the interpretation in the case when $D=3$ where we have $0\leq \beta \leq \frac{\pi}{2}$ and $0 \leq \frac{r}{L} \leq \frac{\pi}{2}$. We visualize the geometry in Fig. \ref{pic:GeomInter}. The two extreme cases are the quarter-circles $QP$ and $QP'$. To generate the one parameter family of surfaces we lift the semi-circle $QQ'$ with Q and $Q'$ fixed and move P along the green circle up until to $P'$. Then $r_{\star}$ corresponds to the position of $P$ along the green circle when it is moving towards $P'$. An example $QP_{1}$ is given in blue. Since both of them are quarter-circles with the same radius, they have the same amount of length.
This tells us that the existence of the one parameter family of entangling surfaces highly depends on the geometry of dS. So if the geometry is deformed by matter, this existence will not be natural anymore.

\section{General Warpfactor}
\label{general}

In order to understand the structure of the EE in more detail, we may consider turning on sources in the holographic dual to slightly deform the spacetime. In this case we end up with metrics of the general form \eqref{ds} but with a more general warpfactor $A(r)$. We want to insist that
\begin{itemize}
\item $A(r)$ still asymptotes to horizons at $r_\text{min}$ and $r_\text{max}$ that are indistinguishable from the ones appearing in dS$_d$ sliced DS$_D$ (or dS$_d$ sliced AdS$_{D}$ for that matter). 
\item For the sake of convenience, we will consider symmetric deformations but our final conclusion is independent of this assumption, as will be indicated later.
\item $A(r)$ has a maximum at some $r_m$ between $r_\text{min}$ and $r_\text{max}$, so $A'(r_m)=0$ with $A''(r_m)<0$. Without loss of generality we can continue to chose $r_\text{m}=0$,  but we should allow that $r_\text{max}/L$ and $r_\text{min}/L=-r_\text{max}/L$ deviates from $ \pi/2$ and $-\pi/2$, respectively, due to extra stretching of space due to the backreaction.
\end{itemize}

In this case we still get a localized graviton around $r=0$ and we can declare this to be the locus at which the $d$ dimensional holographic dS$_d$ dual lives. The curvature radius of this dS$_d$ slice is
\beq
L_* = e^{A(0)}
\eeq
The Newton constants are related by \eqref{genericnewton}. The Lagrangians for class U and class D RT surfaces read
 \beq
{\cal L}_{U} =  \cos^{D-3}(\beta) e^{(D-3) A}  \sqrt{1 + e^{2A}
 (\beta')^2}.
\eeq
and
\beq
{\cal L}_{D} =  \cos^{D-3}(\beta) e^{(D-3) A} \sqrt{(r')^2 + e^{2A}},
\eeq
respectively. Interestingly one can see that both the special class U solution
\beq \rho(r) = 1 \eeq
and the generic class D solution
\beq r(\rho) = 0 \eeq
still solve the equations of motion for general $A$. The former only relied on the fact that at the horizon on the patch $\cos'(\beta)$ vanishes, latter relied on the fact that $A'(0)=0$. The class U surface is a horizon in the bulk, the entanglement entropy associated with its area is given by
\beq
\label{horizonhuggingee}
S = \frac{A}{4 G} = \frac{ A_{D-3} \int_{-r_\text{max}}^{r_\text{max}} e^{(D-3) A}dr }{4 G} = \frac{ A_{D-3}  }{4 g} = S_{dS}.
\eeq
This entangling surface {\it always} reproduces the correct dS$_d$ entropy - both the RT surface and Newton's constant pick up the same factor of $\int e^{(D-3)A}$. Even in the presence of generic deformation we find that the dS$_d$ entropy within the DS/dS correspondence is always completely accounted for by the entanglement entropy across the dS$_d$ horizon. As emphasized in the introduction, this is consistent with the general results of \cite{Emparan:2006ni,Myers:2013lva}.

The situation is quite different for the class D surfaces. Even in the global case where we take the entire spatial volume and trace over one CFT, the class D surface now will have no obvious relation to the DS$_D$ entropy. The volume is only sensitive to $e^{A(0)}$, the value of the warpfactor at its maximum (which sets the curvature radius $L_*$ of the dS$_d$ slice and hence its volume). The Newton constant on the other hand is given by the integral of $A$. So generically the entanglement entropy will be different from the dS$_d$ entropy.

If one were to assume that the dS$_d$ entropy provides an upper bound on any entanglement entropy that can be obtained for a density matrix obtained from tracing, one would expect constraints on $A(r)$, and hence on the allowed matter, that ensure that the class D entanglement entropy is never larger than the dS$_d$ entropy. For now we take the class D entanglement entropy the one associated with the global separation, accounting for the full dS entropy before deformation, or equivalently the two sided static patch class D surface, that is the $r^* \rightarrow 0$ limit of our one parameter family of surfaces we found above. As we argued above, each of these surfaces is presumed to be a valid entanglement entropy of some density matrix obtained by a well defined tracing procedure and so one may expect that it should not yield an entropy larger than that of the dS$_d$ entropy on the slice. The analog of \eqref{globalee} becomes
\beq
\label{globalee2}
S_{D,\text{global}} =  \frac{2 (L_*)^{D-2} V_{D-2}}{4 G} = \frac{A_{D-3} L_*^{D-2}}{4 G} \int_{-\pi/2}^{\pi/2} d\beta \, \cos^{D-3} \beta,
\eeq
whereas the dS$_d$ entropy is given by \eqref{horizonhuggingee} above. Consistency hence seems to require
\beq
 1 \geq \frac{S_{D,\text{global}}}{S_{dS}} = \frac{ e^{(D-2) A(0)} \int_{- \pi/2}^{\pi/2} d\beta \, \cos^{D-3} \beta}{\int_{-r_\text{max}}^{r_\text{max}} e^{(D-3) A}dr}.\label{bound}
 \eeq
At first sight, it looks to be hard to extract information from this bound as it compares a local quantity, the maximum warpfactor $e^{A(0)}$, to an integrated quantity, $\int e^{(D-3)A}$. Fortunately we will see in the following subsections that we can make some quite general statements about the bound.

\subsection{Scalar Field in Warped de-Sitter}\label{scalars}
In order to understand the bound eq. \eqref{bound} better, we first consider a scalar field in the warped dS$_D$ geometry eq. \eqref{ds} (with $D\ge3$) given by the action
\beq
        S=\frac{1}{2\,\kappa^2}\int d^{D}x\sqrt{-g}\left(R-2\Lambda\right)-\int d^{D}x\sqrt{-g}\left(\partial_{a}\phi\partial^a\phi+V(\phi)\right),
\eeq
where the corresponding equations of motion can be found in appendix \ref{app1}. For simplicity we give the argument for symmetric warpfactors, where the full integral is simply twice the integral from the middle slice to $r_\text{max}$. For asymmetric warpfactors we can easily get the same results by adding two different contributions from the left and right half of space. Our calculations below in this case can be trivially performed for the left and right half separately. Using eq. \eqref{apeq}, we can write the denominator of eq. \eqref{bound} as
\begin{equation}
        \int_{-r_\text{max}}^{r_\text{max}}e^{(D-3)A}dr=2\int_{-\infty}^{A(0)}e^{(D-3)A}\frac{dA}{A'}=2\int_{-\infty}^{A(0)}\frac{e^{(D-3)A}}{\sqrt{H(r)-\frac{1}{L^2}+\frac{e^{-2A}}{L^2}}}\,dA,
        \label{eq:CA}
\end{equation}
where we defined
\begin{equation}
H(r)=\frac{\kappa^2}{(D-1)(D-2)}\,\left(\phi'^{2}-2\,V(\phi)\right).\label{auxF}
\end{equation}
Notice that this change of integration variable is valid because \eqref{potfree} tells us that:
\begin{equation}
    A''(r)\leq0
\end{equation}
which means that $A'(r)$ is monotonically decreasing. Since $A'$ vanishes at $0$, we have:
\begin{equation}
        A'(0)^{2}=H(0)-\frac{1}{L^2}+\frac{e^{-2A(0)}}{L^2}=0.\label{eq:Zero}
\end{equation}
which tells us that
\begin{equation}
    \begin{split}
    A'(r)>0 \text{ for }-r_\text{max}<r<0, \quad \text{ and } \quad  A'(r)<0\text{ for } 0<r<r_\text{max}.
    \end{split}
    \label{eq:Monot}
\end{equation}
This equips us with the monotonicity property of $A(r)$ which allows the change of variable in \eqref{eq:CA} from $r$ to $A$. This does not depend on the assumption that the deformation is symmetric.
By shifting variables, $A(r)=\Delta(r)+A(0)$, we find
\begin{equation}
\begin{split}
      \int_{-r_\text{max}}^{r_\text{max}}e^{(D-3)A}dr= 2\,L\, e^{(D-2)A(0)}\int_{-\infty}^{0}\frac{e^{(D-3)\Delta}\,d\Delta}{\sqrt{e^{-2\Delta}-\frac{L^2 H(r)-1}{L^2 H(0)-1}}} \leq2\, L\, e^{(D-2)A(0)}\int_{-\infty}^{0}\frac{e^{(D-3)\,\Delta}d\Delta}{\sqrt{e^{-2\Delta}-1}}. 
      \end{split}\label{eq:Aint}
\end{equation}
The inequality holds because of the following argument. Using the equations of motion we find
\begin{equation}
       H'(r)=-\frac{2\,\kappa^2}{D-2}A'\phi'^{2}.
\end{equation}
With this the monotonicity property of $A(r)$ eq. \eqref{eq:Monot} tells us that
\begin{equation}
        H(0)\leq H(r),\text{for } r\in[-r_\text{max},r_\text{max}].
\end{equation}
Using \eqref{eq:Zero}, we get:
\begin{equation}
    H(0)<\frac{1}{L^2},
\end{equation}
and hence
\begin{equation}
        \frac{1-L^2\, H(r)}{1-L^2\, H(0)}\leq1.
\end{equation}
Solving the integral in eq. \eqref{eq:Aint} we see  that
\begin{equation}
 \int_{-r_\text{max}}^{r_\text{max}}e^{(D-3)A}dr\le \frac{ e^{(D-2)A(0)}\,\sqrt{\pi}\,\Gamma(\frac{D}{2}-1)}{\Gamma(\frac{D-1}{2})}.
\end{equation}
The integral in the numerator of eq. \eqref{bound} is given in eq. \eqref{dsEE} and we immediately see that for scalar fields the bound eq. \eqref{bound} is always violated or just saturated! An example for scalar field which violates the bound is given in appendix \ref{appbound}.
\subsection{Towards the Null Energy Condition}
The independence of our argument in section \ref{scalars} on the precise form of the potential of the scalar field, motivates us to generalize our argument by phrasing it in terms of energy conditions.  The analogue of eq. \eqref{auxF} (in $D\ge3$) reads
\begin{equation}
        H(r)=\frac{2 \kappa^2 L^2}{(D-1)(D-2)}T^{r}_{r}.
\end{equation}
Using the equations of motion, we can write for the derivative
\begin{equation}
   \partial_r T^{r}_{r}=(D-1)A' \kappa^2\,\left(T^{t}_{t}-T^{r}_{r}\right) .
\end{equation}
We can use the same argument for the inequality in case of
\begin{equation}
        -T^{t}_{t}+T^{r}_{r}>0.
\end{equation}
This equips us with the same monotonicity properties as before which imply that the bound is violated. But this is exactly the null energy condition (NEC), obeyed by any reasonable matter. So the bound is in tension with the NEC. Again we emphasize that this does not rely on the assumption of symmetric deformations.

\section{Interpretation}
\label{conclude}

Let us take stock of where we are. We set out to derive swampland bounds on potential matter fields in a de-Sitter (dS) background by comparing two different entropies. For one we took the global class D entanglement entropy $S_{D,\text{global}}$. In the field theory this was interpreted as the entropy associated with the entanglement of left and right CFTs in the entire $r=0$ spatial UV-slice dS$_d$ geometry. If the bulk geometry is DS$_D$, this global L/R entanglement accounts for the entire DS$_D$ entropy, which in this case is equal to the DS$_d$ entropy. The second entropy we looked at was the de Sitter entropy $S_{dS}$ associated with de Sitter gravity on the $d=D-1$ dimensional UV slice. If we follow the standard logic that the maximum entropy of any quantum theory on dS$_d$ is given by the dS entropy we can postulate a bound on allowed matter, which can lead to deformations away from DS$_D$ and hence changes in the entropy. The bound \eqref{bound} required
\beq
\label{boundagain}
\frac{S_{D,\text{global}}}{S_{dS} } \leq 1.
\eeq
Somewhat surprisingly, we found that any reasonable matter (obeying the null energy condition) violates the bound! There are several ways one could interpret this result. For one, our framework may not be consistent. Maybe dS quantum gravity just is never well defined in the presence of matter, or maybe the DS/dS framework is not the correct description. But, more optimistically, one may also expect that the requirement \eqref{boundagain} was simply too strong. $S_{D,\text{global}}$ is the entanglement on the entire spatial volume, which is twice as large as the dS static patch volume. $S_{dS}$ is the maximum entropy available to a single observer on the UV slice dS$_d$. Clearly what we are seeing is that a single dS$_d$ observer has no access to sufficient information to fully describe the higher dimensional DS$_D$ geometry. We require some information from beyond the horizon. While this is somewhat surprising, it may not necessarily be an inconsistency but rather another fascinating property of the DS/dS correspondence.

Moving forward, one would think that a minimum consistency requirement that one would want to impose under any circumstance is
\beq
\label{futurebound}
\frac{S_{D,\text{static}}}{S_{dS} } \leq 1 \quad \Leftrightarrow \quad 
\frac{S_{D,\text{global}}}{S_{dS} } \leq 2.
\eeq
This time we are demanding that the entanglement entropy between L and R CFT entirely within the static patch has to be less than the dS$_d$ entropy; this seems to be a basic consistency requirement on any quantum theory in de Sitter.
We leave a quantitative analysis of what matter would be ruled out by this slightly less stringent bound for the future.

\section*{Acknowledgements}

We are grateful to Eva Silverstein for helpful discussions and encouragements.
This work was supported in part by the U.S.~Department of Energy under Grant No.~DE-SC0011637.
HG is very grateful to his parents and recommenders. SG gratefully acknowledges financial support by the DAAD (German Academic Exchange Service) for a \textit{Jahresstipendium f\"ur Doktorandinnen und Doktoranden}.

\appendix
\section{Equations of motion}\label{app1}
The equation of motion for the scalar field and warp factor in warped dS$_D$ read
\begin{align}
       & \phi''(r)+(D-1)\,A'(r)\,\phi'(r)-V'(\phi)=0\label{eq:scalareom}\\
&\label{eq:Einsteineq1}L^2 \,A'^{\,2}-e^{-2A}+1=\frac{2 \kappa^2\, L^2}{(D-1)(D-2)}\left(\frac{1}{2}\phi'^{2}-V(\phi)\right)\\
\label{apeq}& -A''-\frac{(D-1)}{2}A'^{2}+e^{-2A}\frac{(D-3)}{2L^2}-\frac{(D-1)}{2L^2}=\frac{\kappa^2}{D-2}\,\left(\frac{1}{2}\phi'^{2}+V(\phi)\right),
\end{align}
where $\kappa$ is the coupling of the matter to gravity.
Since the equations of motion for the warp factor are linearly dependent, we can build linear combinations, i.e. eliminate the scalar potential
\begin{equation}
        -A''(D-2)-\frac{D-2}{L^2}e^{-2A}=\kappa^2\,\phi'^{2},\label{potfree}
\end{equation}
and write down an equation for the potential
\begin{equation}\label{eq:pot}
-\frac{(D-2)}{2\kappa^2} \left ( A''+(D-1)A'^{\, 2}-\frac{D-2}{L^2}\, e^{-2A}+\frac{D-1}{L^2} \right ) = V(\phi).
\end{equation}

\section{Example: The bound for a linear scalar field}\label{appbound}
\begin{figure}[h]
    \centering
    \includegraphics[width=7cm]{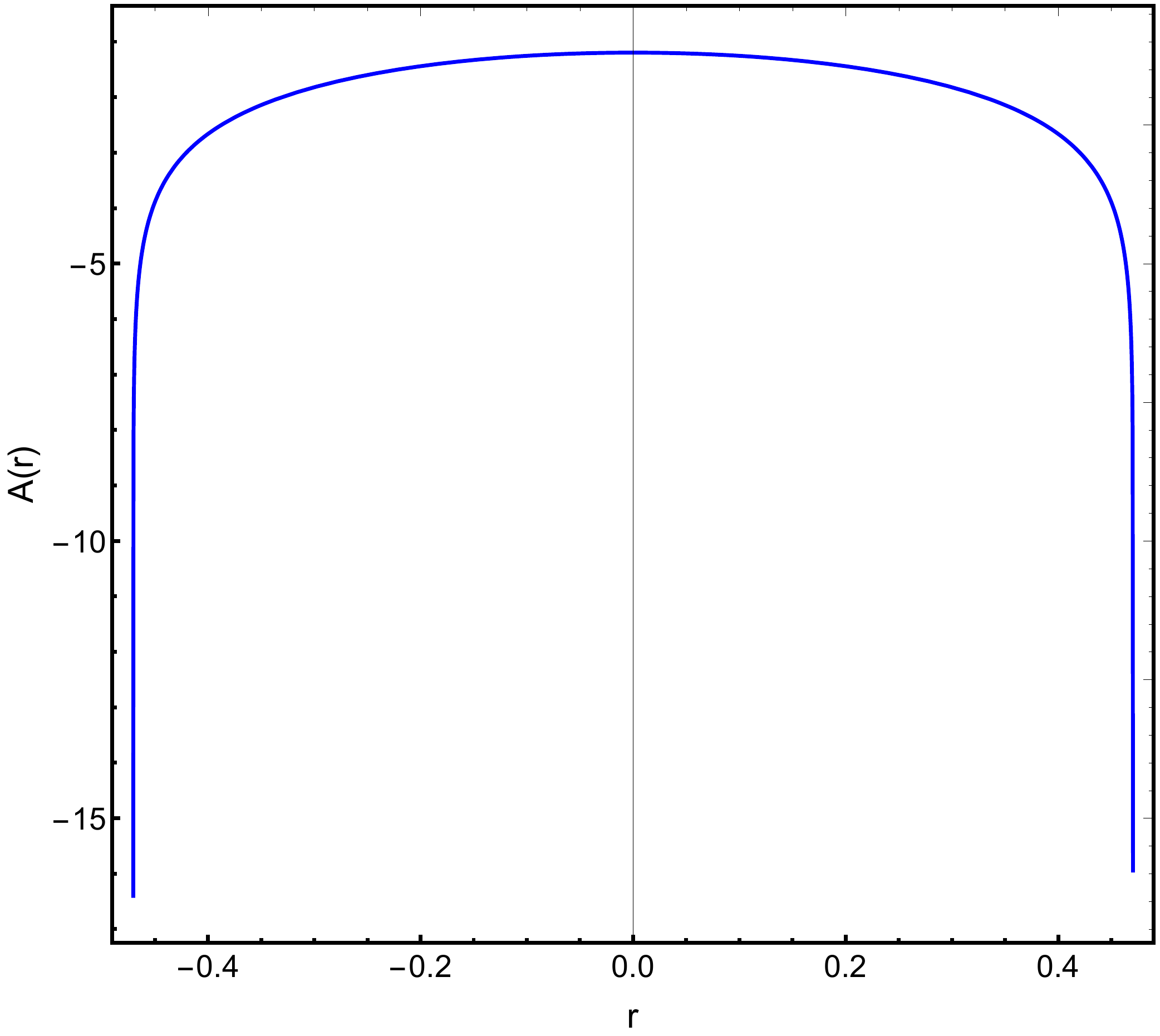}\quad  \includegraphics[width=7cm]{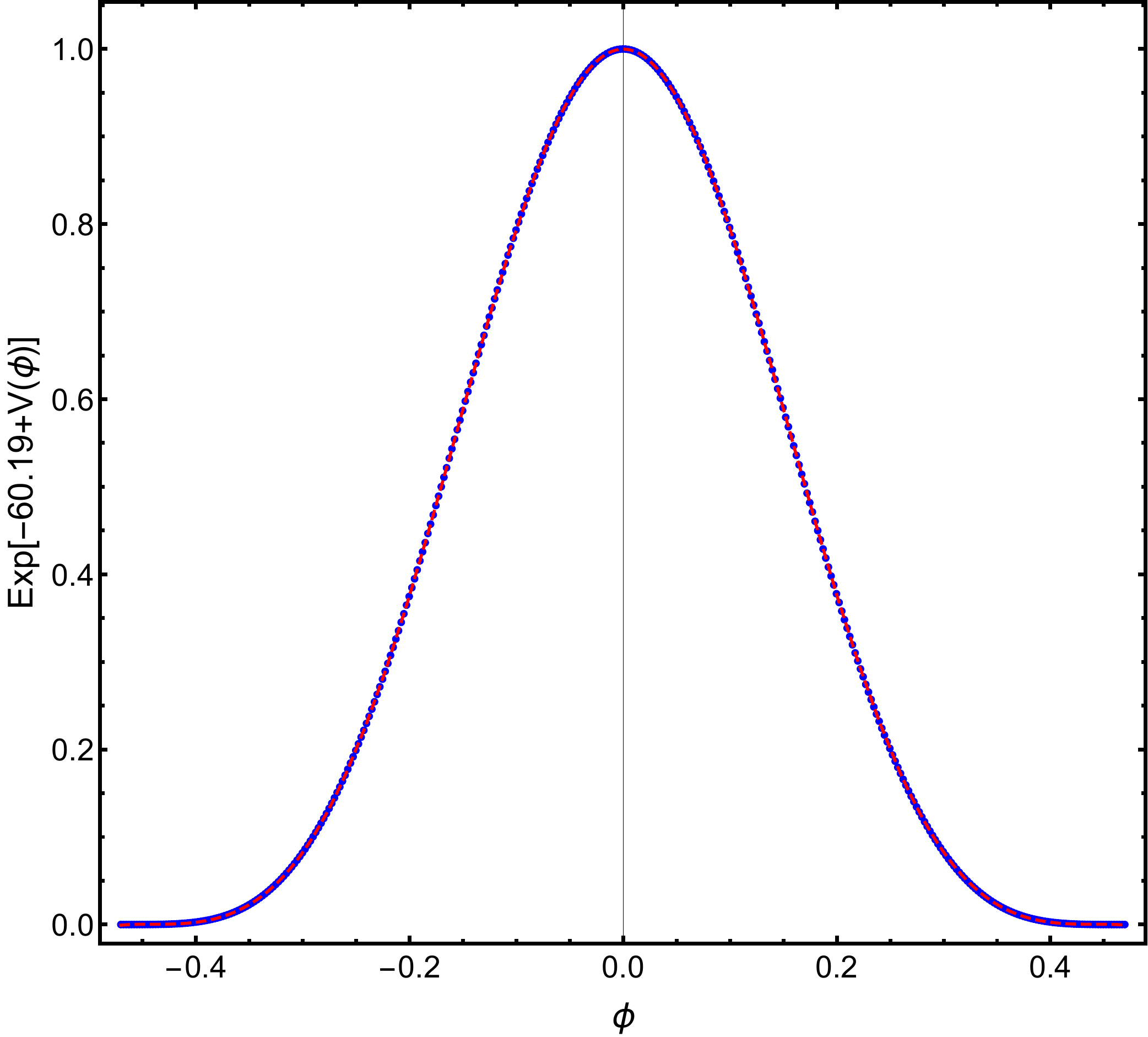}
        \caption{\textbf{Left:} Warp factor $A(r)$ in presence of a linear scalar. The warp factor $A$ diverges at $r_\text{min}$ and $r_\text{max}$. \textbf{Right:} The blue line depicts the reconstructed potential of the scalar field. The dashed red line depicts the fit with an 10th order polynomial $\text{exp}(-60.19+V(\phi))=1.00
-8.12\, \phi^2 + 
 75.53\, \phi^4 - 
 367.12\, \phi^6 + 
 940.42\, \phi^8 - 
 1016.03\, \phi^{10}$ .}\label{pic:A+V}
\end{figure}

To illustrate the bound, we consider a linear scalar field $\phi(r)=c_1 r$ in $D=5$ with $L=\kappa=c_1=1$. Solving eq. \eqref{potfree}, subject to the boundary condition that $A(r)$ behaves asymptotically as eq. \eqref{asympdS} we find may find $A(r)$ as depicted in 
figure \ref{pic:A+V}. Plugging $A(r)$ in eq. \eqref{bound} leads to an entanglement entropy ratio of
  \begin{equation}\frac{S_{D,\text{global}}}{S_{dS}}=1.013\ge 1,
  \end{equation}
which violates the bound. In order to write down a Lagrangian for the scalar field, we have to reconstruct the potential. This may be done by solving the equation for the potential \eqref{eq:pot} and is depicted in figure \ref{pic:A+V}.

\newpage
\bibliographystyle{JHEP}
\bibliography{dsdsee}
\end{document}